\documentclass{aa}

\usepackage{graphicx}
\usepackage{txfonts}
\usepackage{color}

\begin{document}

\title{Lopsided galactic disks in IllustrisTNG}

\author{Ewa L. {\L}okas
}

\institute{Nicolaus Copernicus Astronomical Center, Polish Academy of Sciences,
Bartycka 18, 00-716 Warsaw, Poland\\
\email{lokas@camk.edu.pl}}


\abstract{
A significant fraction of nearby late-type galaxies are lopsided. We study the asymmetry of the stellar component in a
sample of well-resolved disky galaxies selected from the last snapshot of the Illustris TNG100 simulation based on
their flatness and rotational support. Among 1912 disks, we identify 161 objects with significant asymmetry in terms of
the $m=1$ Fourier mode of the stellar component within (1-2) stellar half-mass radii and describe their properties
using three representative examples. The profiles of the $m=1$ mode typically increase with radius, and the
corresponding phase is constant in the asymmetric region, signifying a global distortion. Following the evolution of
the lopsided disks over time, we find that their history is rather uneventful and the occurrence of the
asymmetry is fairly recent. Only about 1/3 of the lopsided disks experienced any strong interaction recently that could
have led to the distortion of their shape: 24\% were affected by a more massive object and 9\% underwent a gas-rich
merger. Still, a majority of lopsided disks show a significant increase in their recent star formation rate. The most
frequent mechanism for the formation of lopsided disks thus seems to be asymmetric star formation probably related to
gas accretion, although the distortions in the gas and stars are not strongly correlated. This picture is supported by
the finding that the lopsided population on average contains more gas, has higher star formation rate, lower
metallicity and bluer color than the remaining disks. These correlations are similar to those seen in real galaxies,
even though the fraction of simulated lopsided disks (8\%) is much lower than in observations (30\%). The observed
correlation between the presence of the asymmetry and a bar is not reproduced either. These discrepancies may be due to
overquenching or insufficient resolution of IllustrisTNG simulations. }


\maketitle

\section{Introduction}

Many galactic disks in the nearby Universe appear to be non-axisymmetric \citep{Jog2009}. The
feature was first systematically studied in \citet{Baldwin1980}, which considered lopsided galaxies as a
subclass of spirals. They examined about 20 best known galaxies with asymmetries and proposed that the asymmetry is
caused by a lopsided pattern of elliptical orbits. \citet{Rix1995} quantified the asymmetry by odd Fourier modes and
found that out of 18 face-on spirals considered, about 1/3 were substantially lopsided. These findings were confirmed
and extended to 60 field spiral galaxies by \citet{Zaritsky1997}, 54 early-type disk galaxies by \citet{Rudnick1998},
147 galaxies of the OSUBGS sample by \citet{Bournaud2005} and 167 galaxies of different luminosities and morphologies
by \citet{Zaritsky2013}. Galaxies with lopsided morphology are expected to also show large-scale asymmetries in their
kinematics; for example, in the form of different shapes of the rotation curve on both sides of the galaxy
\citep{Swaters1999, Noordermeer2001, Jog2002, Eymeren2011a, Ghosh2021}.

A few observational studies, including \citet{Zaritsky1997}, \citet{Conselice2000}, and \citet{Rudnick2000}, have
suggested a link between the presence of the lopsided disk and recent star formation events and an excess of blue color
in the galaxy. This connection was later established in \citet{Reichard2009}, which studied lopsidedness in a
sample of $\sim$ 25,000 nearby galaxies from the Sloan Digital Sky Survey (SDSS). They found a strong correlation
between lopsidedness of the galactic disk and the youth of the galaxy stellar population. The lopsided galaxies from
SDSS turned out to be more star-forming, more metal-poor, and younger than the symmetric objects. These correlations
were later confirmed for other galaxy samples by \citet{Wang2011} and \citet{Yesuf2021} and are consistent with
scenarios that deliver lower metallicity gas into the galaxy central region.

The possible origin of lopsidedness in galactic disks has been addressed in studies using simulations of
galaxy evolution \citep{Zaritsky1997, Bournaud2005, Mapelli2008, Ghosh2021}. Most often, interactions between
galaxies, such as mergers and flybys, were considered as plausible mechanisms for the generation of such distortions.
It soon became clear, however, that they cannot explain all occurrences of this phenomenon given its presence in
isolated galaxies as well. One other possibility is that a lopsided disk forms as a result of a perturbations from a
lopsided halo \citep{Jog1997, Jog1999, Levine1998}. It has also been shown that long-lived lopsided global modes in the
stellar component can exist in a galaxy evolving in isolation \citep{Saha2007, Dury2008}.

\begin{figure*}
\centering
\includegraphics[width=17cm]{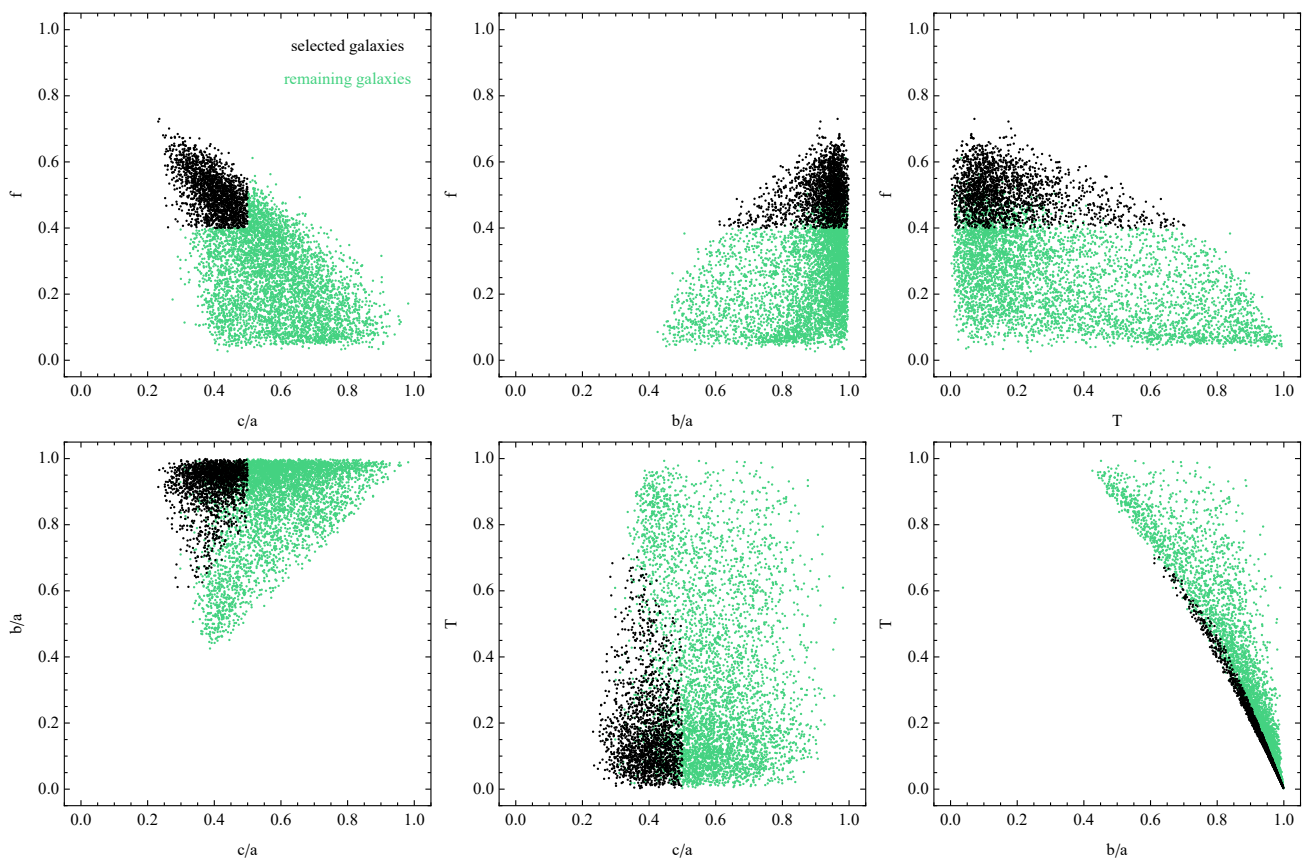}
\caption{Selection of disk galaxy sample. The six panels show the properties of the selected
galaxies in different parameter combinations: the axis ratios $b/a$ and $c/a$, the triaxiality parameter $T,$ and
the rotation parameter $f$. Black points correspond to 1912 selected disk galaxies, and the green points show the
remaining 4595 galaxies of the total of 6507 well-resolved objects.}
\label{selection}
\end{figure*}

In light of the observationally detected correlations, the particularly promising scenario for the formation of the
asymmetric shape in a spiral galaxy seems to be the one proposed by \citet{Bournaud2005}. Their simulations
demonstrated that galaxy interactions and mergers can trigger strong lopsidedness, but to explain all the observational
results, it is required that in many cases lopsidedness results from cosmological, asymmetric accretion of gas on
galactic disks. This picture is confirmed by studies of individual isolated lopsided spiral galaxies such as P11695
\citep{Vulcani2018}.

Recently, it has become possible to investigate the origin of asymmetry in galactic disks not only by the means of
controlled simulations, as was done before, but also in the cosmological context. New sets of cosmological simulations
now available are able to produce large samples of galaxies with sufficient resolution to study their morphology. In
this work and for this purpose, we used the simulations of galaxy formation from the IllustrisTNG project
\citep{Springel2018, Marinacci2018, Naiman2018, Nelson2018, Pillepich2018}. The simulations follow the
evolution of galaxies from the early Universe to the present by solving gravity and hydrodynamics, and
applying additional prescriptions for star formation, galactic winds, magnetic fields, and the feedback from black holes.
Various studies performed thus far have demonstrated that these simulations are able to reproduce many of the observed
properties of galaxies, including their morphologies \citep{Nelson2018, Genel2018, Rodriguez2019}. The set of
simulations comprises the results obtained with different resolution in boxes of 300, 100, and 50 Mpc on one side
(referred to as TNG300, TNG100, and TNG50, respectively).

Asymmetries in galaxies have been addressed using IllustrisTNG in \citet{Watts2020}, which focused on the
distortions in the gas using HI spectral lines in TNG100 galaxies, and \citet{Whitney2021}, which considered
asymmetry in TNG300 and TNG50 galaxies in the context of mergers. Here, we studied the asymmetry of the stellar
component of late-type galaxies in the TNG100 simulation, which provides a sufficient sample of galaxies with
good resolution. In Section~2, we describe our selected sample of galaxies and the identified subsample of
lopsided disks. In Section~3, we discuss the basic properties of these asymmetric objects using three representative
examples. Section~4 is devoted to the origin of the lopsided shape in simulated galaxies and the discussion follows in
Section~5.

\section{Sample selection}

For this study, we made use of the publicly available simulation data from the IllustrisTNG project as described by
\citet{Nelson2019}. We chose the TNG100 run, that is, the simulation performed in the 100 Mpc box, which contains a
sufficient number of galaxies with different morphologies. In order to have enough resolution
in each object and thus obtain a sample eligible for morphological analysis, we selected the galaxies at $z=0$ by
restricting the sample of subhalos to those with the total stellar masses greater than $10^{10}$ M$_\odot$, which
corresponds to about $10^4$ stellar particles per object. This criterion is fulfilled by 6507 objects in the final
snapshot of the Illustris TNG100 simulation. In order to select disk galaxies among them, we imposed two additional
conditions: we required the galaxies to be rotationally supported and rather thin.

Following \citet{Joshi2020}, we assume that disk galaxies have the rotation parameter $f > 0.4$. The rotation
parameter is defined as the fractional mass of all stars with circularity parameter $\epsilon > 0.7$, where
$\epsilon=J_z/J(E),$ and $J_z$ is the specific angular momentum of the star along the angular momentum of the galaxy,
while $J(E)$ is the maximum angular momentum of the stellar particles at positions between 50 before and 50 after the
particle in question in a list where the stellar particles are sorted by their binding energy \citep{Genel2015}.

The disk galaxies were supposed to be sufficiently thin if their shortest-to-longest axis ratio $c/a$ of the
stellar component was lower than 0.5. For these values, we used (and reproduced) the measurements based on the mass
tensor of $c/a$ within two stellar half-mass radii, $2 r_{1/2}$, provided by the Illustris team in the Supplementary
Data Catalogs of stellar circularities, angular momenta, and axis ratios and calculated as described in
\citet{Genel2015}. The axis ratios were estimated from the eigenvalues of the mass tensor of the stellar mass obtained
by aligning each galaxy with its principal axes and calculating three components ($i =$ 1, 2, 3): $M_i = (\Sigma_j m_j
r^2_{j,i}/\Sigma_j m_j)^{1/2}$, where $j$ enumerates over stellar particles, $r_{j,i}$ is the distance of stellar
particle $j$ in the $i$-axis from the center of the galaxy, and $m_j$ is its mass. The eigenvalues were sorted so that
$M_1 < M_2 < M_3,$ which means that the shortest-to-longest axis ratio is $c/a = M_1/M_3$, while the
intermediate-to-longest axis ratio is $b/a = M_2/M_3$.

We note that the values of $c/a$ estimated from the mass tensor within $2 r_{1/2}$ are not directly comparable to axis
ratios estimated from the whole optical images of galaxies, which are usually much lower. For example, a realistic
$N$-body realization of a Milky Way-like galaxy has the $c/a$ ratio within two disk scale lengths on the order of 0.2
in spite of quite a flat appearance \citep{Lokas2019}. However, even the flattest galaxies formed in
IllustrisTNG have $c/a > 0.2$, which means that they are generally thicker than the observed population of disks,
probably as a result of limited resolution \citep{Haslbauer2022}. It may therefore seem more proper to call them disky
galaxies, but in the following we refer to them as disks for simplicity.

The sample of disk galaxies with the rotation parameter $f > 0.4$ and the shortest to longest axis ratio $c/a < 0.5$
contains 1912 objects. The properties of the selected galaxies in comparison to the whole sample are shown in
Fig.~\ref{selection}. In the six panels of the figure, we plot the positions of the galaxies in different planes of
parameters: the axis ratios $b/a$, $c/a$, the triaxiality parameter $T = [1-(b/a)^2]/[1-(c/a)^2]$, and the rotation
parameter $f$. In the three upper panels of Fig.~\ref{selection}, the rotation parameter $f$ lies in the vertical axis.
These panels illustrate the selection with the simple cutoff at $f > 0.4$. In the lower three panels, the positions of
the selected galaxies are less obvious, and in particular we find that they all lie along the border of the whole
distribution in the $T - b/a$ plane in the lower right panel of the figure. We also note that many of the disks are
quite triaxial with $T$ reaching the values of 0.7, but none is decidedly prolate ($T > 0.7$).

\begin{figure}
\centering
\includegraphics[width=7.5cm]{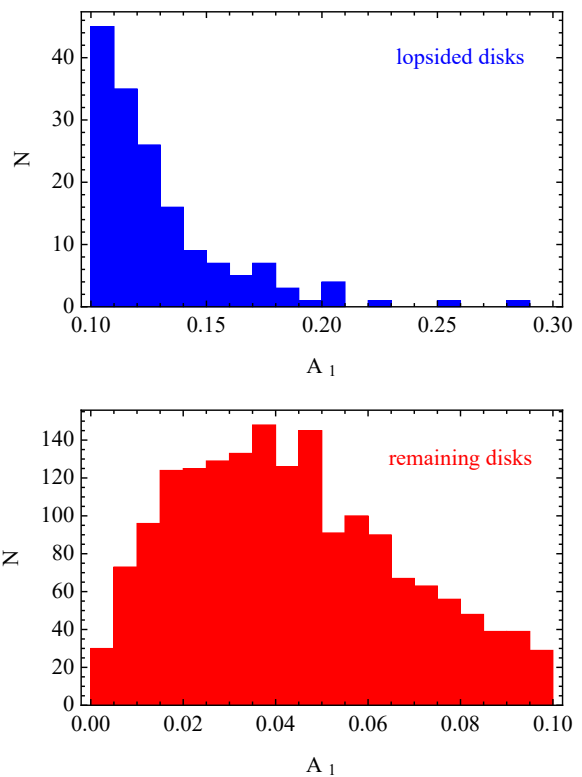}
\caption{Distributions of $m = 1$ Fourier mode values $A_1$ for lopsided disks with $A_1 > 0.1$ (upper panel) and
remaining disks with $A_1 < 0.1$ (lower panel). Measurements of $A_1$ were done within $(1-2) r_{1/2}$. The bin
size is 0.01 in the upper panel and 0.005 in the lower one.
}
\label{histogramsa1}
\end{figure}

In order to quantify the degree of lopsidedness in all 1912 disks at the present time, we calculated different modes of
the Fourier decomposition of the surface density distribution of stellar particles projected along the short axis, $A_m
(R) = | \Sigma_j m_j \exp(i m \theta_j) |/\Sigma_j m_j$, where $\theta_j$ is the azimuthal angle of the $j$th star,
$m_j$ is its mass, and the sum goes up to the number of particles in a given radial bin along the cylindrical radius
$R$. The galaxies were centered on their dynamical center, which for IllustrisTNG subhalos is determined as the position
of the particle with the minimum gravitational potential energy.

In observational studies of lopsidedness, the measurements of the Fourier modes are usually performed in the radial
range of $(1.5 - 2.5) R_{\rm e}$, where $R_{\rm e}$ is the galaxy exponential radius \citep{Rix1995, Bournaud2005}. In
order to make the comparison with observations meaningful we performed the measurements in a similar radial bin. Since
the disks of the simulated galaxies are only approximately exponential, the stellar half-mass radius, $r_{1/2}$, is a
better, more robust measure of the galaxy size, used for estimating different properties of IllustrisTNG galaxies. In
addition, for an exponential disk $1.5 R_{\rm e}$ contains almost half the light (to be exact, $r_{1/2} = 1.68
R_{\rm e}$). Therefore we chose to estimate the Fourier modes from stars with cylindrical radii $R$ in the
$(1-2) r_{1/2}$ range.

The values of the $m = 1$ mode, $A_1$, provide a measure of the
asymmetry of the stellar distribution. Among the sample of 1912 disks, we identified 161 galaxies with significant
asymmetry $(A_1 > 0.1$) in this outer range, and in the following we refer to these galaxies as lopsided
disks. The distribution of the $A_1$ values for the 161 lopsided disks and the remaining galaxies with $A_1 < 0.1$ are
shown in Fig.~\ref{histogramsa1}. We see that most of the disks have very low values of the asymmetry, while in the
lopsided sample the values of $A_1$ reach 0.29, although only seven galaxies have $A_1 > 0.2$. The mean $A_1$
for the whole sample of 1912 galaxies is 0.051 and the median 0.044.

\section{Properties of lopsided disks}

\begin{table*}
\caption{Properties of three selected lopsided disks from IllustrisTNG at $z=0$.
The masses are the total masses of different components
and the $g-r$ color was estimated from all stars. The values of $A_1$ were measured in the range $(1-2)
r_{1/2}$ and the rest of the parameters within $2 r_{1/2}$.
}
\label{properties}
\centering
\begin{tabular}{c c r r c c c c c c c c c}
\hline\hline
ID \     &  $M_{\rm stars}$      & $M_{\rm gas}$ \ \ \ \ & $M_{\rm dm}$ \ \ \ \ &$r_{1/2}$& $b/a$  & $c/a$ & $T$  & $A_1$ & $f_{\rm gas}$ & SFR                   & $g-r$ & $Z$  \\
         & [$10^{10}$ M$_\odot$] & [$10^{10}$ M$_\odot$] & [$10^{11}$ M$_\odot$]& [kpc]   &        &       &      &       &               & [M$_\odot$ yr$^{-1}$] & [mag] & $[Z_\odot]$ \\ \hline
222275   &  1.52                 &  1.61 \ \ \ \         &  2.06 \ \ \ \        & 3.17    &  0.92  &  0.42 & 0.18 & 0.165 & 0.27          & 2.67                  &  0.25 & 1.56        \\
436552   &  1.39                 &  3.15 \ \ \ \         &  2.35 \ \ \ \        & 3.82    &  0.87  &  0.43 & 0.30 & 0.289 & 0.22          & 2.63                  &  0.26 & 1.65        \\
568873   &  1.27                 &  5.21 \ \ \ \         &  3.50 \ \ \ \        & 4.86    &  0.90  &  0.45 & 0.24 & 0.260 & 0.33          & 1.44                  &  0.33 & 1.42        \\
\hline
\end{tabular}
\end{table*}

\begin{figure}
\centering
\includegraphics[width=4.4cm]{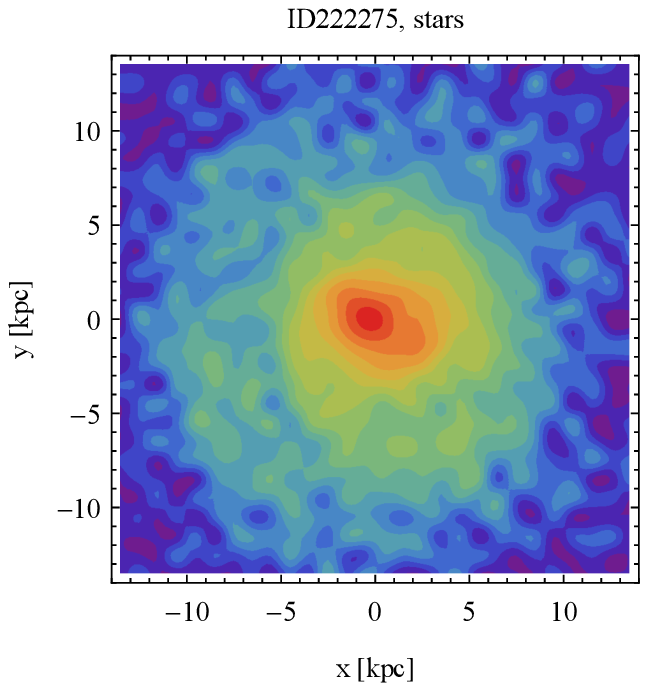}
\includegraphics[width=4.4cm]{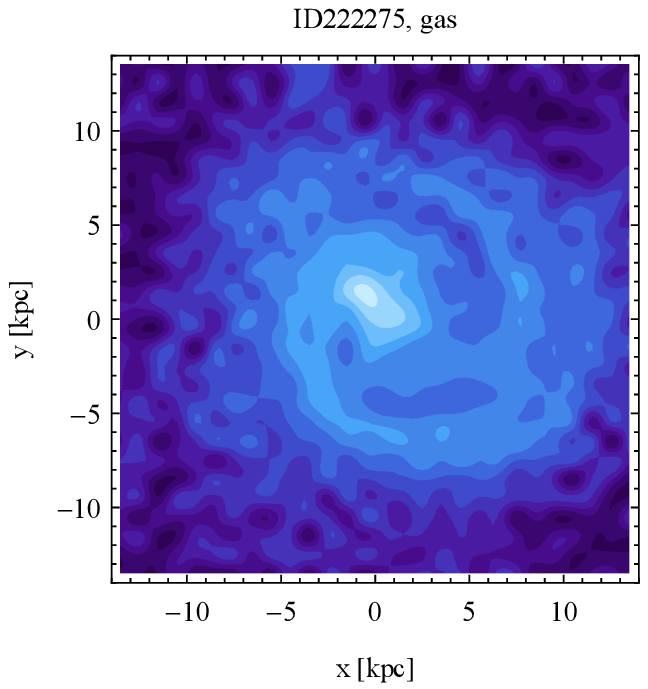} \\
\vspace{0.3cm}
\includegraphics[width=4.4cm]{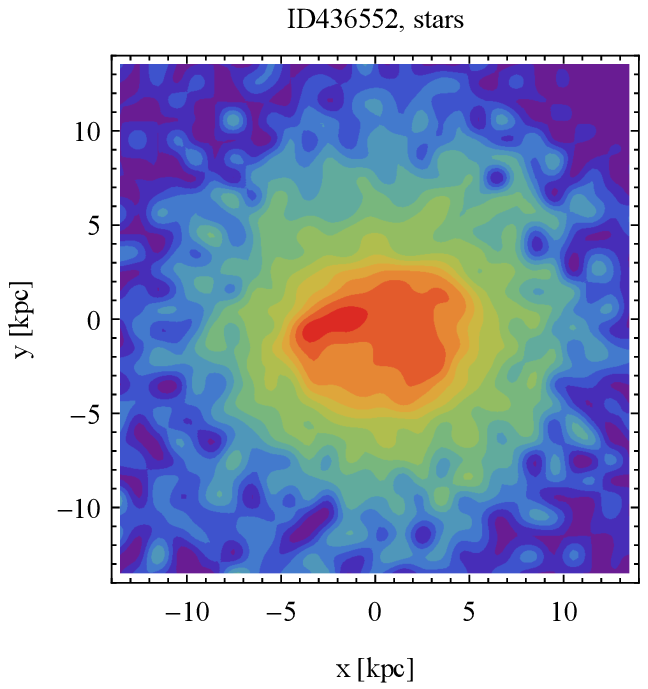}
\includegraphics[width=4.4cm]{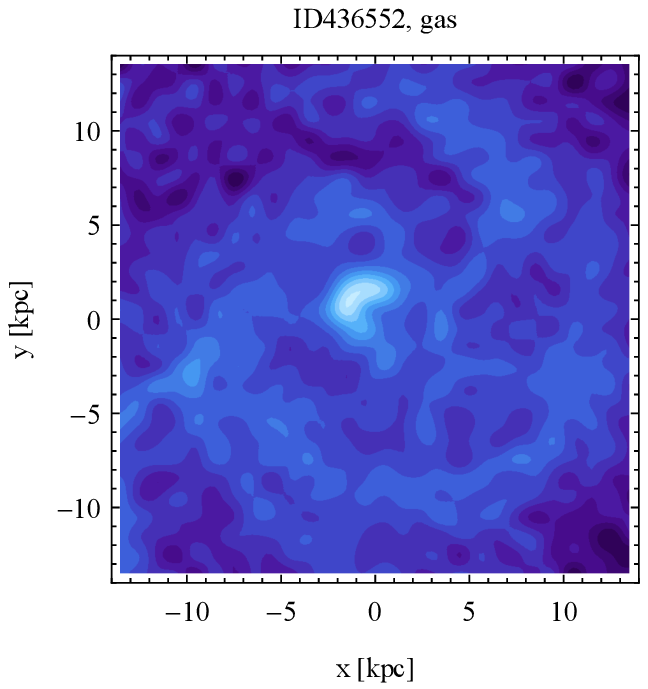} \\
\vspace{0.3cm}
\includegraphics[width=4.4cm]{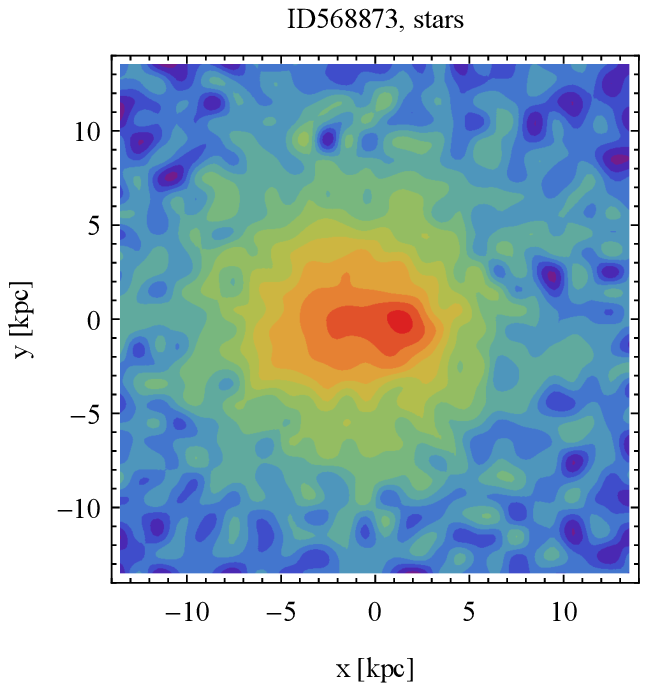}
\includegraphics[width=4.4cm]{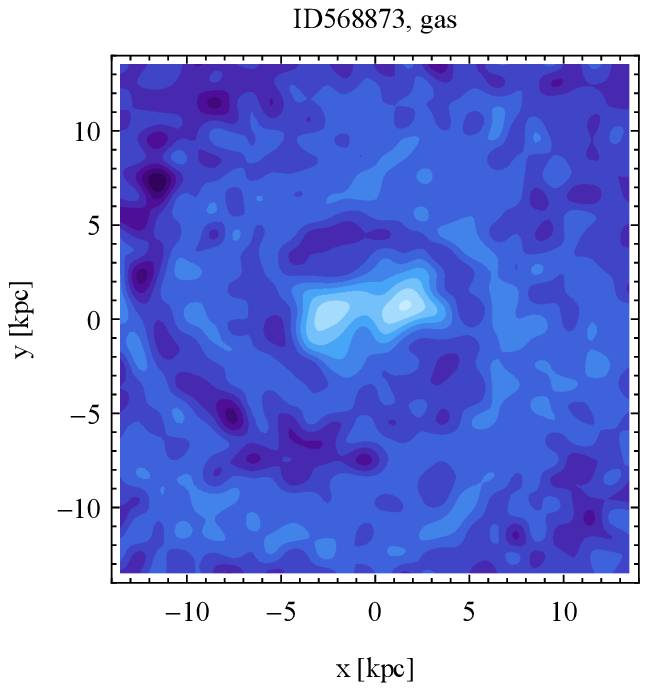} \\
\caption{Surface density distributions of stars (left column) and gas (right column) in the face-on view at the present
time for three selected lopsided disks from IllustrisTNG (from top to bottom). The surface densities are
normalized to the central value in each case and the contours are equally spaced in $\log \Sigma$.}
\label{surden}
\end{figure}

In this section, we describe the properties of the sample of 161 lopsided disks in more detail, focusing on three
representative examples with high values of $A_1$ measured within $(1-2)
r_{1/2}$. The surface density distributions of these three galaxies in the face-on view at the present time
corresponding to the last simulation snapshot ($z=0$) are plotted in Fig.~\ref{surden}. The left column panels show the
stellar component and the right column ones show the gas. The asymmetry of the stellar disks is well visible in the images
and so is the asymmetry in the gas, although the latter has a different form. The gaseous disks are usually more
extended and less uniform, taking the form of rings and spirals. The basic properties of the three galaxies at the
present time are given in Table~\ref{properties}. The first column of the table gives the identification number of the
subhalo in the IllustrisTNG catalog and the next three list the total masses of the stars, gas, and dark matter. The
fifth column gives the stellar half-mass radii, $r_{1/2}$, and the next three columns list the shape parameters: $b/a$,
$c/a,$ and $T$. The values of $A_1$ are provided in the ninth column. The remaining columns list the values of gas
fraction, star formation rate, color, and metallicity.

\begin{figure}
\centering
\includegraphics[width=7cm]{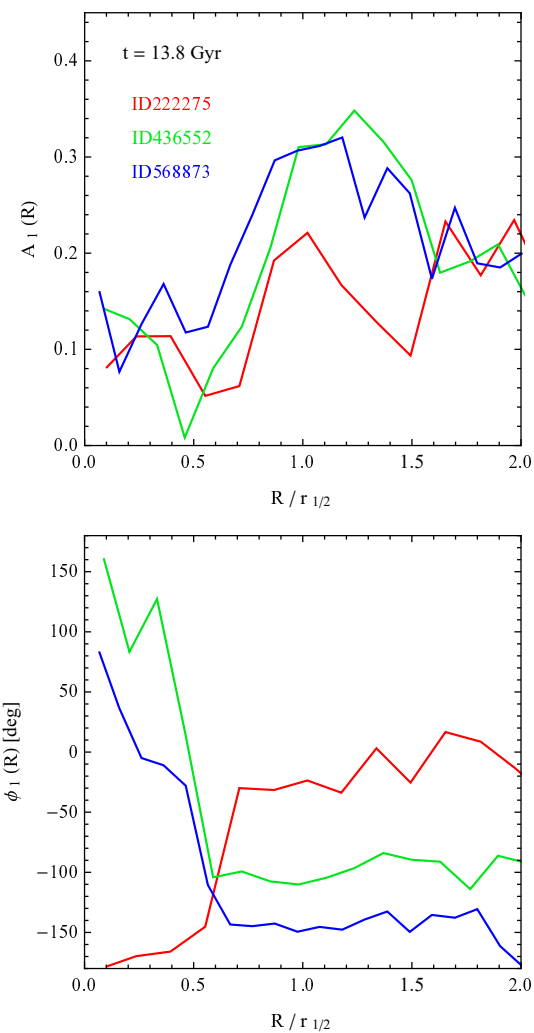}
\caption{Profiles of $m = 1$ Fourier mode $A_1 (R)$ (upper panel) and its phase angle $\phi_1 (R)$ (lower panel)
at the present time for three selected lopsided disks from IllustrisTNG. Measurements were carried out in bins of
$\Delta R = 0.5$ kpc of cylindrical radius $R$ in the face-on projection.}
\label{a1phaseprofiles}
\end{figure}

\begin{figure}
\centering
\includegraphics[width=7.5cm]{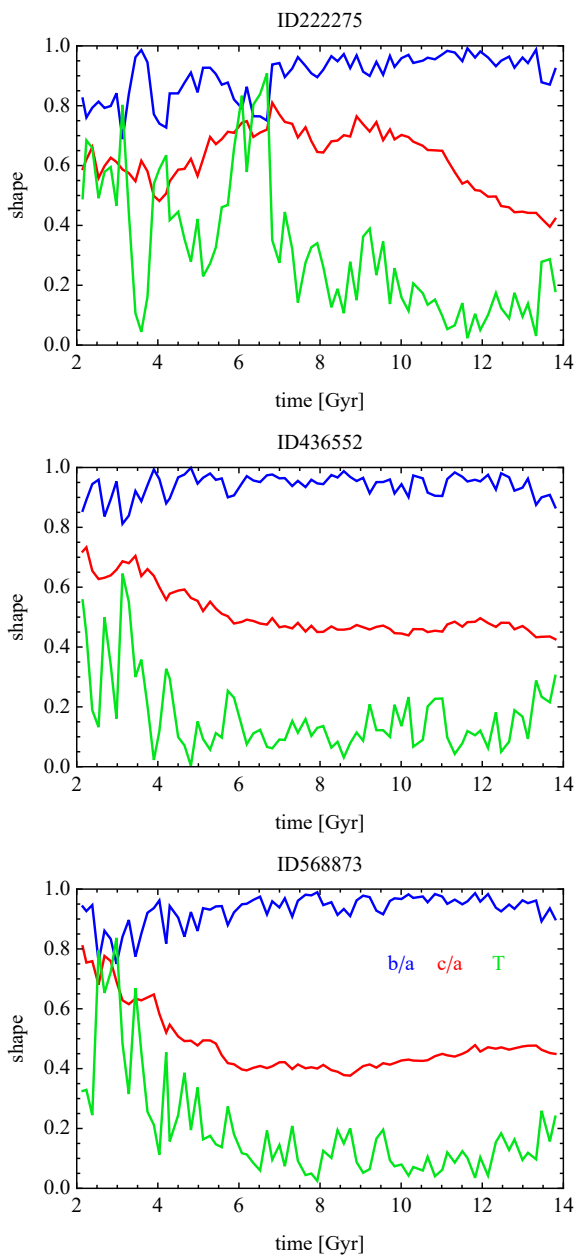}
\caption{Evolution of shape parameters for three selected lopsided disks from IllustrisTNG. The blue, red and green
lines show, respectively, the axis ratios $b/a$, $c/a$ and the triaxiality parameter $T$. Measurements were done within
$2 r_{1/2}$.}
\label{shape_all}
\end{figure}

\begin{figure}
\centering
\includegraphics[width=7.6cm]{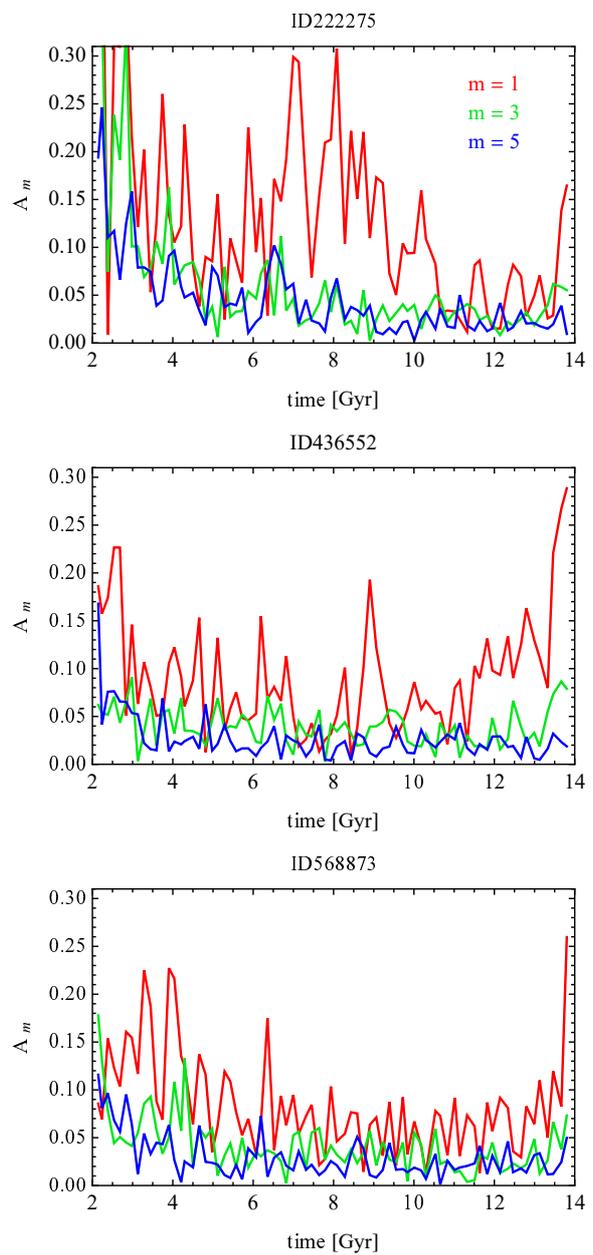}
\caption{Evolution of odd Fourier modes $A_1$, $A_3,$ and $A_5$ for three selected lopsided disks from IllustrisTNG.
Measurements were done in the range of $(1-2) r_{1/2}$.}
\label{oddmodestime_all}
\end{figure}

\begin{figure}
\centering
\includegraphics[width=7.5cm]{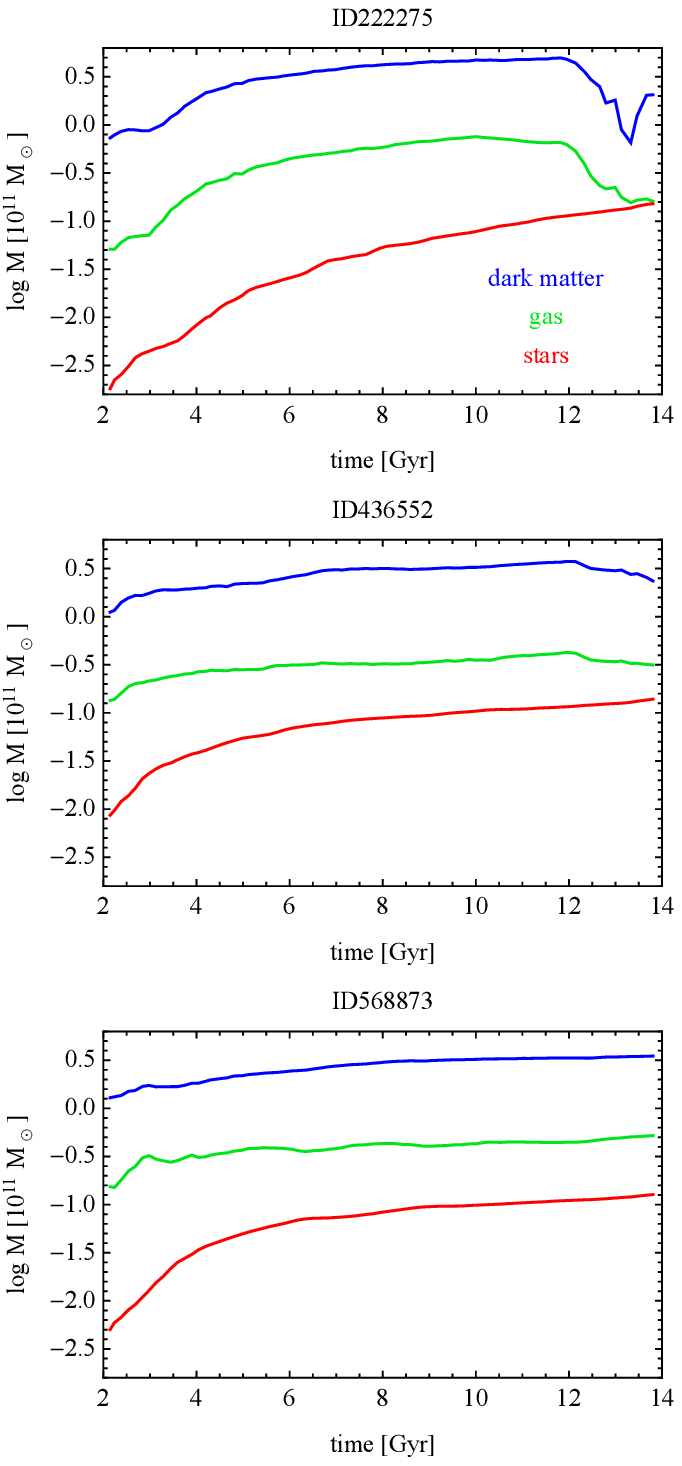}
\caption{Evolution of total mass in different components for the three selected lopsided disks from
IllustrisTNG. The red, green, and blue lines show, respectively, the stellar, gas, and dark matter masses.}
\label{mass_all}
\end{figure}

In order to quantify the asymmetry in more detail for each of the 161 lopsided galaxies, we calculated the profiles of
the $A_1$ mode and their phase angles as a function of the cylindrical radius $R$ in the face-on projection. Three
examples of such profiles for the selected lopsided galaxies are shown in Fig.~\ref{a1phaseprofiles} as a
function of $R/r_{1/2}$. Interestingly, the shapes of the $A_1$ profiles (the upper panel of
Fig.~\ref{a1phaseprofiles}) are similar in the three cases: they have low values of $A_1$ on the order of 0.1
near the center, which increase with radius reaching a maximum of about 0.2-0.35, to finally decrease again
down to about 0.2 at the outer radii. The peak of the $A_1$ profile is reflected in the behavior of the phase (the
lower panel of Fig.~\ref{a1phaseprofiles}) in the sense that the phase remains constant in the range of radii where the
peak occurs. This means that $A_1$ is a global mode, extending over a substantial fraction of the disk: $(0.5 - 2)
R/r_{1/2}$. We find that most of the 161 lopsided galaxies have similar shapes of the $A_1 (R)$ and $\phi_1 (R)$
profiles, which means that they seem to be general properties of lopsided disks.

\begin{figure}
\centering
\includegraphics[width=7.5cm]{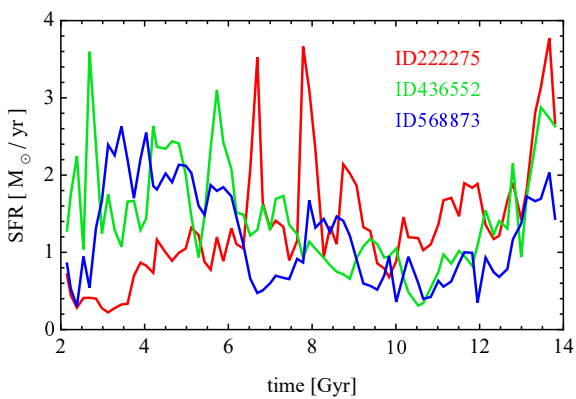}
\caption{Evolution of star formation rate within $2 r_{1/2}$ for the three selected lopsided disks from
IllustrisTNG.}
\label{sfr}
\end{figure}

The first measurements of the $A_1$ profiles in real galaxies were performed in \citet{Rix1995}, which
discovered that they grow in the outer parts of galaxies. \citet{Rudnick1998} averaged the profiles of $A_1$ modes for
lopsided galaxies among their early-type disk sample and found them to be increasing with radius in a way similar to
the ones shown in the upper panel of Fig.~\ref{a1phaseprofiles}. \citet{Angiras2006} and \citet{Angiras2007} measured
the profiles of $A_1$ in HI surface density maps of galaxies in the Eridanus and Ursa Major groups, respectively, while
\citet{Eymeren2011b} performed such measurements for 70 galaxies from the Westerbork HI Survey of Spiral and Irregular
Galaxies. It turns out that in many cases their $A_1$ profiles are growing with radius and similar to those presented
here, although the variety of profile shapes among the observed galaxies is much larger.

In addition to the properties in the final simulation output, which we used to define our sample, we also looked at
the evolution of the lopsided galaxies in time. Measurements of the axis ratios $b/a$, $c/a,$ and the triaxiality
parameter $T$ (within $2 r_{1/2}$) reveal that, except for their present lopsidedness, all these galaxies are bona fide
disks that preserved their disky morphology for a long time. Three examples of the evolution of these shape parameters
are shown in Fig.~\ref{shape_all} for our selected lopsided disks. Their history seems quite uneventful, with high $b/a$
and low $T$ values characteristic of oblate systems, preserved for a long time until the present. Significant
departures from this quiet evolution can only be seen for ID222275 around $t = 6$ Gyr when this galaxy experienced two
mergers and was temporarily distorted. The evolution of these properties is similarly simple for the rest of
our lopsided disks, except for those with bars which have lower $b/a$ and higher $T$ resulting from the prolate
component.

In spite of their long and quiet existence as disks, the lopsided galaxies still must have acquired their asymmetry at
some point in their history. It is thus interesting to check the evolution of the odd Fourier modes in time.
Examples of these are plotted in Fig.~\ref{oddmodestime_all}, where in addition to $A_1$ we also include the higher
modes $A_3$ and $A_5$. Interestingly, the present high values of $A_1$ are often a very recent occurrence,
appearing only in the last one or two simulation outputs, although they are sometimes (as in the case of ID436552)
preceded by a longer period of enhanced or oscillating $A_1$, but at a lower level. The strong peaks of $A_1$ were more
frequent in the past, because then the forming galaxies were more often undergoing mergers and distorted as a result.
Notable examples of such events are the peaks of $A_1$ at $t = (7-8)$ Gyr for ID222275 (the upper panel of
Fig.~\ref{oddmodestime_all}) corresponding to the strong variation of the shape visible in Fig.~\ref{shape_all},
resulting from mergers, as mentioned above. We note that the $A_3$ and $A_5$ modes are usually subdominant with
respect to $A_1$ so the latter is the most useful and sufficient to describe the departures from symmetry.

\section{Origin of the lopsided shape}

A few possible scenarios have been proposed in the literature regarding the origin of lopsided galactic disks
\citep{Bournaud2005, Mapelli2008, Ghosh2021}. The mechanisms include galactic interactions in the form of mergers and
flybys, ram pressure stripping of the gas, and asymmetric star formation in isolated disks. In this section we try to
discriminate between them and find if there is a prevalent way to form lopsided disks.

It is relatively easy to check if the galaxies belonging to our lopsided sample are affected by tidal interactions with
more massive objects. Such events manifest themselves very clearly in the evolution of the total mass in dark matter
and gas in the IllustrisTNG data. When the galaxy in question passes near a more massive object, its mass is stripped
and assigned to the more massive companion. Such effects can be considered significant if a galaxy loses a
substantial fraction of its maximum mass. We find that for our lopsided galaxy sample, 38 out of 161 objects (24\%)
lost more than 10\% of their dark masses and only ten (6\%) lost more than 50\%. The most affected galaxy lost 80\%
of its dark mass, so the interactions were never strong enough to strip almost all of the galaxy's dark matter as is
the case of objects on tight orbits around a massive galaxy cluster \citep{Lokas2020}.

A good example of this subsample of galaxies is ID222275, which about 1 Gyr ago interacted with a group of
galaxies including two objects of mass of the order of $10^{12}$ M$_\odot$. The mass loss in dark matter and gas in
this galaxy is well visible in the upper panel of Fig.~\ref{mass_all} showing the evolution of total mass in time.
As a result of this interaction, the gas is ram-pressure stripped and its distribution is similar to that of jellyfish
galaxies \citep{Yun2019}. The changed distribution of the gas could have affected the stellar disk and cause its
departure from symmetry. A similar process could be at work in other galaxies of this subsample.

\begin{figure*}
\centering
\includegraphics[width=7.5cm]{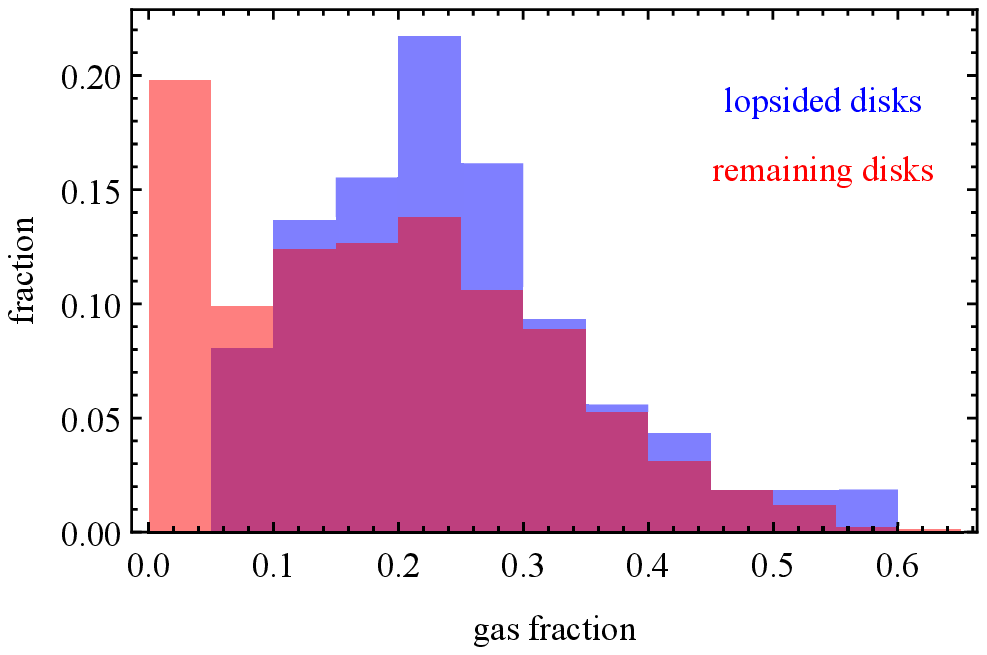}
\hspace{0.5cm}
\includegraphics[width=7.5cm]{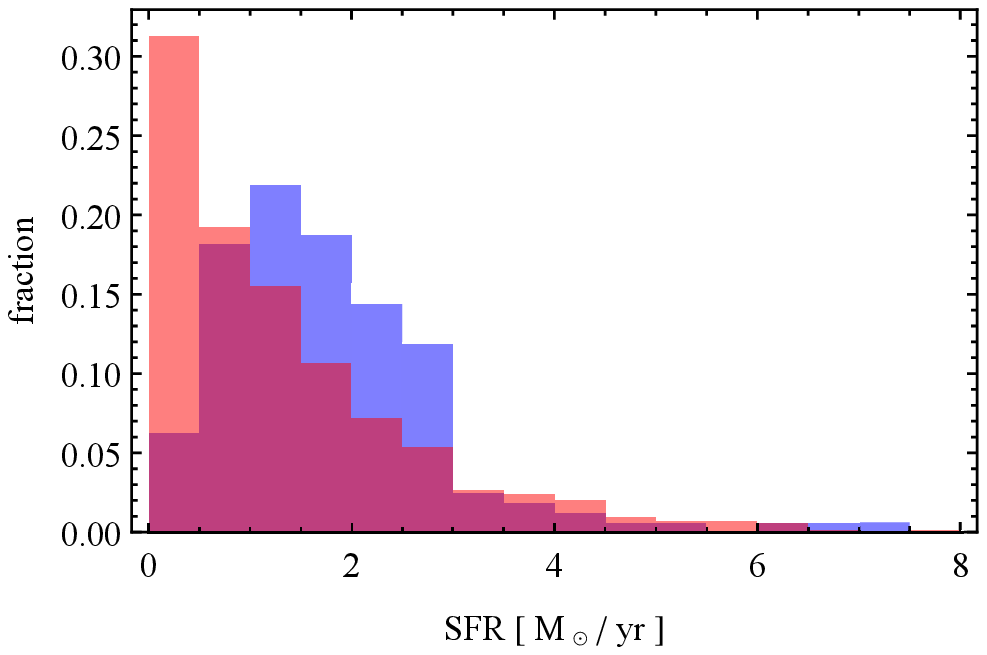}\\
\vspace{0.3cm}
\includegraphics[width=7.5cm]{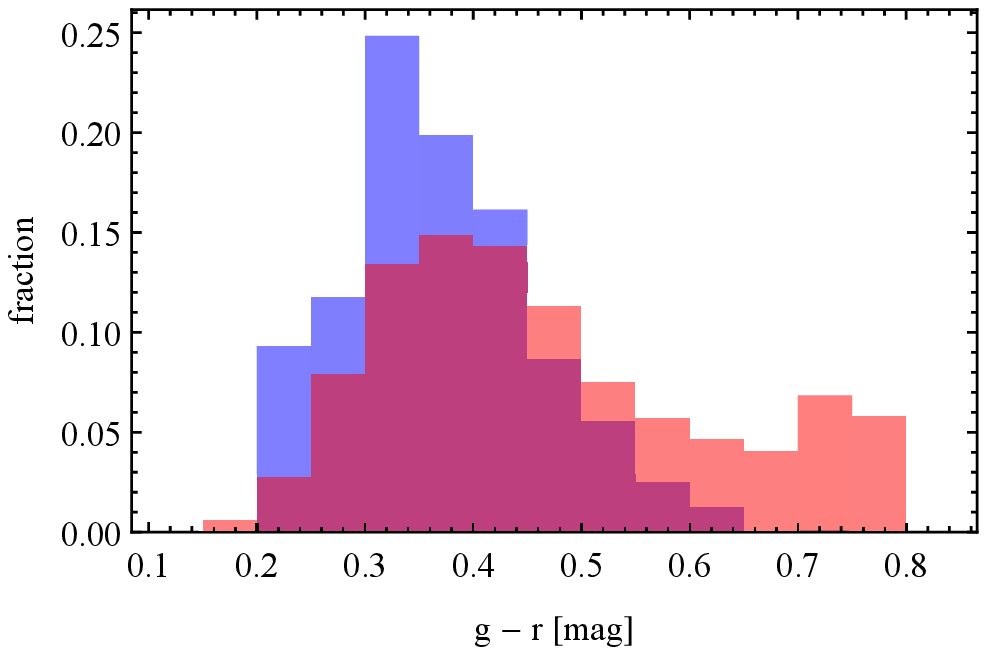}
\hspace{0.5cm}
\includegraphics[width=7.5cm]{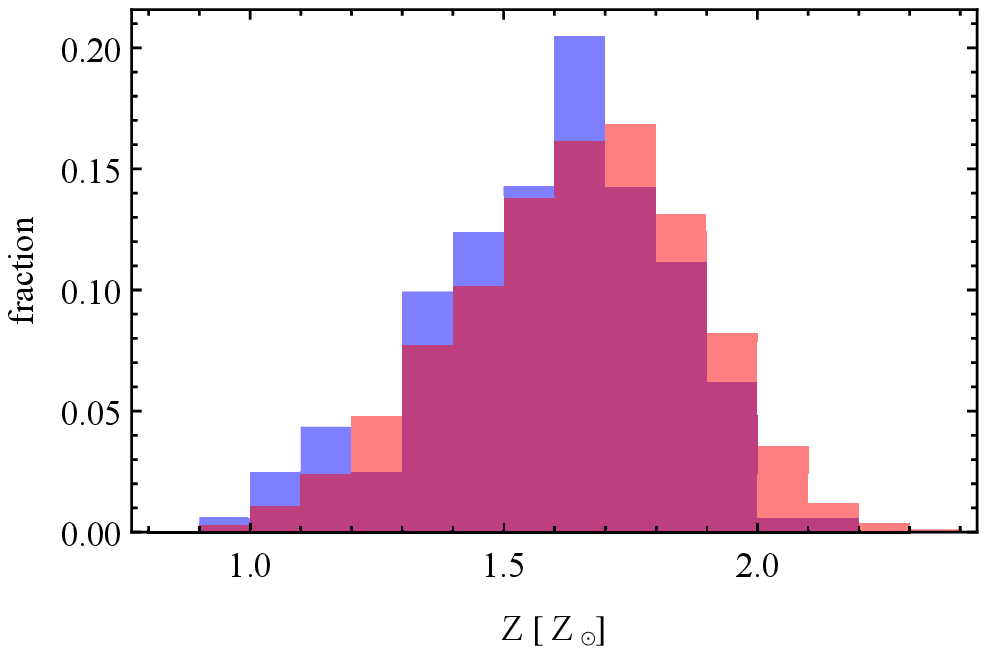}
\caption{Distributions of gas fractions (upper left), star formation rates (upper right), color (lower left),
and metallicity (lower right) for lopsided disks with
$A_1 > 0.1$ (blue) and remaining disks with $A_1 < 0.1$ (red). Measurements of the properties were
done within $2 r_{1/2,}$ except for the color, which is estimated from all stars in the galaxy. All histograms were
normalized to unity.}
\label{histogramsgfsfr}
\end{figure*}

By inspection of the merger trees of the lopsided disks available in the simulation data, we found that only 15
out of 161 (9\%) galaxies experienced a significant gas-rich merger in their recent history ($z < 0.1$), although
almost all continue to accrete small dark subhalos with mass of the order of $10^{8}$ M$_\odot$ until the
present epoch. It is extremely difficult to ascertain to what extent these more significant mergers could be the cause of
the present lopsidedness of the disks.

An example of galaxies belonging to this subsample is ID436552. It has experienced a relatively recent ($z \sim 0.1$)
merger with a satellite that had orbited it for a long time with five pericenter passages. The disturbance caused by
the merger may have increased the inner gas content and could cause the asymmetric star formation resulting in the
lopsided stellar disk. However, this galaxy is also approaching a bigger neighbor at present, losing
dark matter and gas in the process; this is shown in the middle panel of Fig.~\ref{mass_all}. Therefore, ram pressure
stripping of the gas in the outer parts could also affect its dynamics and shape.

Another mechanism for generating disk lopsidedness, also related to interactions, relies on a long-lived halo
distortion caused by a tidal encounter \citep{Weinberg1995} and the disk's response to it, which results in disk
lopsidedness \citep{Jog1997}. However, the resolution of the dark matter halo is not sufficient in the IllustrisTNG
simulations to study such subtle effects; that is, there are too few dark particles in the radial range occupied by the
stars and their softening length is quite large.

For the remaining galaxies, we were unable to identify any interactions that could have caused the distortion of the
shape. These are isolated objects still growing in mass and forming stars. Their common feature is that even if their
star formation rates (SFR) are not very high, they all contain a significant amount of gas. Still, about 40\% of
lopsided galaxies show a significant increase in SFR in their recent history. This trend is also visible in the
evolution of SFR for our three selected lopsided disks shown in Fig.~\ref{sfr}. It seems, therefore, that the most
frequent mechanism for the formation of lopsided disks is asymmetric star formation.

The galaxy ID568873 is a good example of this category. As can be seen from the lower panel of Fig.~\ref{mass_all},
its stellar and gas content do not show any abrupt changes in recent history. Although its SFR was not very high in
recent history, its stellar disk has become decidedly lopsided in the final simulation output.

In order to verify that indeed asymmetric star formation is the main cause of lopsidedness, we compared the gas
fractions $f_g = M_{\rm gas}/(M_{\rm gas} + M_{\rm stars})$ and SFRs at the present time for our sample of lopsided
disks and the remaining disks in the simulation. The histograms showing the distributions of these quantities within $2
r_{1/2}$ are plotted in the two upper panels of Fig.~\ref{histogramsgfsfr}. We can see that the distributions for these
two samples of galaxies are very different. There are no gas-free galaxies among the lopsided disks, and their gas
fractions can be as high as 0.6, with the most typical values being 0.2-0.25 (with the median of the distribution equal to
0.23). On the other hand, a significant number of objects are completely devoid of gas among the remaining disks,
and their gas fractions are typically lower (with the median of 0.18). A similar difference is seen in the distribution
of the SFRs. All lopsided disks maintain a non-zero level of star formation (with the median of 1.6 M$_\odot$
yr$^{-1}$), while the remaining disks include many quiescent objects and, on average, have a lower SFR (with the median
of 1.0 M$_\odot$ yr$^{-1}$). These two measures of activity are obviously not independent, since in the IllustrisTNG
simulations the star formation is a derivative of gas density and for all galaxies the gas mass and the SFR within $2
r_{1/2}$ follow each other closely.

The two lower panels of Fig.~\ref{histogramsgfsfr} compare the distributions of the $g-r$ color for the whole galaxy
and the metallicity of stars within $2 r_{1/2}$ for the lopsided and the remaining disks. We can see that the asymmetric
galaxies with the median $g-r$ color of 0.36 are decidedly bluer than the rest of the disk population, which have the
median of 0.43. We note that there is only one lopsided disk with $g-r > 0.6$, which can be considered as a
threshold separating the red from the blue population \citep{Nelson2018}, while among the rest of the disks there are
many red galaxies. The metallicity distributions of the lopsided and symmetric disks are also slightly
different: the asymmetric galaxies typically have lower metallicity with the median of 1.62 $Z_\odot$, while the
remaining galaxies are on average more metal-rich with the median of 1.66 $Z_\odot$. These distributions are
consistent with the picture involving the accretion of low-metallicity gas onto the galaxy as one of the scenarios
leading to lopsidedness.

\begin{figure}
\centering
\includegraphics[width=7cm]{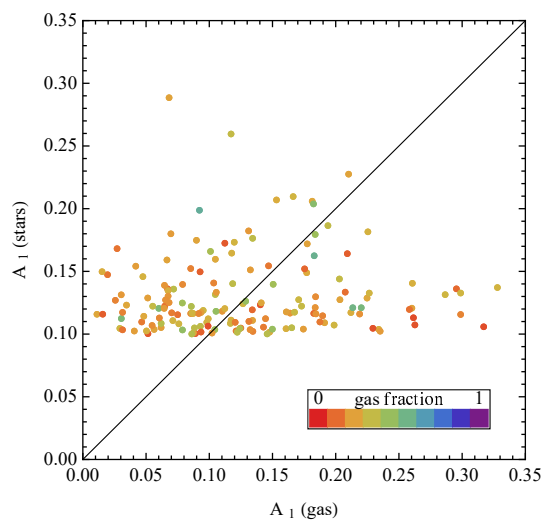}
\caption{Values of $m = 1$ Fourier mode $A_1$ for the stars versus those of the gas within $(1-2) r_{1/2}$ for 161
lopsided galaxies at the present time. The points were color-coded by the gas fraction measured within $2 r_{1/2}$. The
diagonal line indicates the equality of the two quantities.
For clarity, two data points with $A_1 > 0.4$ for the gas were not included in
the plot. }
\label{a1starsgas}
\end{figure}

\section{Discussion}

Using the sample of disk galaxies from the IllustrisTNG project, we calculated the measures of asymmetry in the form of
the $A_1$ Fourier mode of the stellar distribution in the face-on view and identified 161 objects with $A_1 > 0.1$
in the radial range between one and two stellar half-mass radii. These lopsided galaxies are quite similar to
each other in many aspects. They all evolved rather quietly, forming disks early and preserving their oblate shapes for
a long time. In most of them, the value of $A_1$ varies strongly throughout the galaxy's history, especially early
on, reflecting a higher incidence of mergers at that time. In at most 1/3 of the lopsided disks, the asymmetry could
be induced by interactions, either by significant wet mergers or tidal effects and ram pressure stripping caused by more
massive neighbors. The lopsided disks exhibit higher gas fractions and SFRs, bluer colors, and slightly lower
metallicities than the remaining disks, which suggests that asymmetric star formation following the accretion of
low-metallicity gas from the galaxy neighborhood is the dominant mechanism leading to their formation.
We note, however, that this scenario probably cannot explain the asymmetry detected in the old stellar component
\citep{Rix1995, Zaritsky2013}, since the asymmetry in stars is likely to be smeared
out by differential rotation in a few Gyr.

Unfortunately, it is not possible to confirm this scenario more convincingly since the gas cannot be directly traced in
IllustrisTNG simulations. Still, we can verify if the disks are also lopsided in the gas component as this asymmetric
gas could be forming stars and causing the formation of asymmetric stellar disks. For this purpose, we measured
the Fourier mode $A_1$ for the gas in the same region, namely within $(1-2) r_{1/2}$. The values of $A_1$ for the stars
are shown as a function of $A_1$ for the gas in Fig.~\ref{a1starsgas} with the points color-coded by the gas fraction
in the galaxy within $2 r_{1/2}$. The diagonal line indicates the equality between the values for the two components.
We can see that there is little correlation between the $m = 1$ mode for the stars and for the gas, although the points
with a higher gas fraction (green) cluster more around the line. We note that $A_1$  can be much higher for the gas than
for the stars, reaching values as high as 0.66. However, this is the case mostly for galaxies with a low gas fraction
within $2 r_{1/2}$, which means that the number of gas particles included in the calculation is low and the results may
be quite noisy. In fact, the majority of galaxies (88 out of 161, or 55\%) have their $A_1$ for the gas lower than for
the stars. We conclude that there is no direct relation between the present global asymmetry of the gas and the stars.

In general, the gas distribution is more extended and takes the form of rings and spirals, which may contribute randomly
to the measurement of $A_1$. In addition, according to the IllustrisTNG model the stars
are formed from gas with density above some threshold, so the asymmetry in the stars may be caused by stars
forming in a particular overdense region of the gas that has little relation to its global distribution.
Moreover, the present asymmetry of the gas may differ from the one at the time the stars were formed. Measuring
such subtle effects on small subpopulations of stars of different age would, however, require much larger resolution
than is presently available for IllustrisTNG galaxies.

\begin{figure}
\centering
\includegraphics[width=7.5cm]{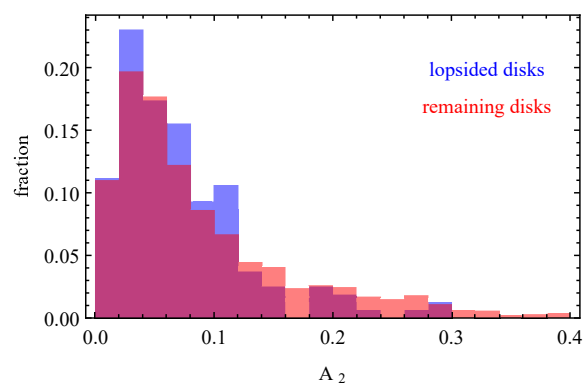}
\caption{Distributions of $m = 2$ Fourier mode values $A_2$ for lopsided disks with $A_1 > 0.1$ (blue) and
remaining disks with $A_1 < 0.1$ (red). Measurements of $A_2$ were done within $2 r_{1/2,}$ and the histograms were
normalized to unity. }
\label{histogramsa2}
\end{figure}

It is also interesting to look at the correlations between the lopsidedness of the stellar disks and the presence of
other morphological features such as bars or spiral arms. Although spiral arms are not well resolved in the Illustris
TNG100 simulation used in this study, bars can be reliably detected using the $m = 2$ Fourier mode and have been
studied in the past \citep{Peschken2019, Rosas2020, Zhou2020, Zhao2020, Lokas2021a, Lokas2021b}. In order to address
this issue, we calculated the bar mode $A_2$ within $2 r_{1/2}$ for the lopsided disks and the remaining ones in our
sample. The distributions of this quantity for the two samples are shown in Fig.~\ref{histogramsa2}, and we can see that
they are quite similar with the median $A_2$ values of 0.059 for the lopsided sample and 0.062 for the remaining disks.
In addition, adopting the threshold of $A_2 > 0.2$ as the value indicating a strong bar, we note that only seven
out of 161 lopsided disks (4\%) are strongly barred, while for the remaining disks this fraction rises to 11\%. This
result is in disagreement with the one of \citet{Bournaud2005}, whose observational sample of galaxies showed that
the presence of lopsidedness is correlated with the presence of bars or spiral arms.

Lopsidedness may occur not only in disks but also in bars, and it may be interesting to consider a possible relation
between the two phenomena. Recently, we identified a few lopsided bars in the IllustrisTNG simulation
\citep{Lokas2021b} that bear some resemblance to the bar in the Large Magellanic Cloud \citep{Marel2001, Jacyszyn2016}.
These objects were found among the bar-like galaxies studied earlier \citep{Lokas2021a}, which have almost the whole
stellar component in the form of a prolate spheroid with a negligible disk. The lopsided bars are
characterized by significant values of odd Fourier modes $A_3$ and $A_5$ with $A_1$ subdominant with respect to them.
This is in contrast with the lopsided disks studied here in which $A_1$ is always the strongest odd mode.
We checked our seven lopsided disks with the strongest bars for asymmetry in the bar and found that none of the
bars is strongly lopsided; so, there are no objects that would possess both a lopsided disk and a
lopsided bar. The galaxy with the strongest bar asymmetry among those (ID523489) has the $A_3$ value within $2
r_{1/2}$ (typically the strongest odd mode in lopsided bars) on the level of 0.06. The only connection between the two
phenomena thus seems to be the possibility that the formation of lopsided bars is preceded by a temporary occurrence of
a lopsided disk, but quite often the time difference between the two is too large to warrant a causal relation.

In comparison with observations, our analysis seems to yield a much smaller fraction of lopsided disks in the whole
population. We identified 161 galaxies out of 1912 disks as lopsided, which only accounts for about 8\%, while in
observations this percentage is estimated to be at the level of 30\% \citep{Jog2009}. We note that these
numbers should be comparable since we made the measurements in a similar radial range as in observations and
applied the threshold of $A_1 > 0.1$ recommended in observational studies \citep{Bournaud2005} as the one to be used to
distinguish lopsided galaxies from the symmetric ones. The estimated mean value of $A_1$ for the whole sample of 1912
disk galaxies from IllustrisTNG is 0.051, while in observational studies this value is closer to 0.1
\citep{Zaritsky1997, Bournaud2005}. The $A_1$ profiles of the simulated objects show similar radial variation to
nearby galaxies, although for most cases they tend to decrease rather than saturate in the outer disk.

In spite of this, we found that the simulations
reproduce the trends found in observations \citep{Reichard2009}, namely that lopsided disks contain more gas, have
higher SFRs, lower metallicity, and bluer colors than the rest of the late-type galaxies.
If lopsidedness is related to the presence of star-forming, young populations, it is possible that the
low fraction of asymmetric galactic disks in IllustrisTNG is caused by the overquenching effect known to exist in
these simulations \citep{Angthopo2021}. If too many galaxies prematurely stop forming stars in IllustrisTNG,
they are also less likely to generate lopsided stellar distributions.
However, we cannot discard the possibility that the correlation between the star formation and lopsidedness can
be explained by the reversed causal relation; namely that the lopsidedness, originating from any mechanism,
affects the disk dynamics and leads to an increased, asymmetric star formation \citep{Jog1997}.

It is also possible that the limited resolution of the simulations does not allow us to reproduce the subtler
effects in the dynamics of lopsided disks adequately. The simulated disks are, for example, significantly thicker than
observed \citep{Haslbauer2022}, which may affect their dynamics, and in particular the disk response to halo
distortion proposed as one of the scenarios for generating lopsidedness. This could also explain the lack of
correlation between the presence of asymmetry and the presence of a bar.

\begin{acknowledgements}
I am grateful to the referee, Chanda Jog, for very useful comments and to the IllustrisTNG team for making
their simulations publicly available.
\end{acknowledgements}

\end{document}